# Composite operators and algebra constraints: a formalism for highly interacting systems

## Ferdinando Mancini

*Dipartimento di Fisica "E.R. Caianiello" - Unità di Ricerca INFM di Salerno*
*Università degli Studi di Salerno, I-84081 Baronissi (SA), Italy*

**Abstract.** A formalism for the study of highly interacting electronic systems is presented. The proposed scheme is based on two key concepts: composite operators and algebra constraints. Composite field operators, that naturally appear as a consequence of interaction, are promoted to the rank of basic fields in terms of which a perturbation formulation is set up. The formalism is based on the use of Green's function and equation of motion method. The use of composite operators requires a revisitation of the Green's function formulation, where the representation is determined by means of algebra constraints which are a manifestation at macroscopic level of the algebra rules and symmetry properties obeyed at microscopic level.

## 1. INTRODUCTION

Since the discovery of superconductivity it has become more and more clear that the physics of many-particle interacting systems is very rich and complex. The development of technology, the possibility of changing the external thermodynamical parameters up to very extreme regions, the discovery of new materials, have led to an enormous progress in Condensed Matter Physics. New and unsuspected properties have been discovered and we are faced with a revolution whose influence is not limited to the scientific world but is involving all fields. Still, we are touching the top of an iceberg whose dimensions are not clear. The progress in technology and in experimental science has been accompanied by a parallel progress in the development of new theories and new schemes of calculations.

In the last twenty years most of the progresses have been made just in the discovery of new materials with unusual properties. It is believed that the origin of such anomalous behaviors is generally due to strong electronic correlations in narrow conduction bands [1]. In this line of thinking many analytical methods have been developed for the study of highly correlated electron systems [2], such as the Hubbard approximation [3], the spectral density approach [4,5], the non crossing approximation [6-8], the slave boson method [9-11], the $d_\infty$-method [12-16], the projection operator method [17-22], the composite operator method [23-29]. The main difficulties are connected with the absence of any obvious small parameter in the strong coupling regime and with the simultaneous presence of itinerant (spatial correlations) and atomic (pronounced on-site quantum fluctuations) features. According to this, it is extremely difficult to judge the reliability of the results obtained by the various approximation methods. The comparison with the recently accumulated results of numerical simulations, although severely restricted in cluster size and temperature and, therefore, with generally poor momentum and energy resolutions [30], is a unavoidable basic step. The numerical results are certainly a guide for the construction of any microscopic theory and, in any case, the different theoretical formulations should refer to them as experimental results obtained on model Hamiltonians instead of materials.

## 2. COMPOSITE OPERATORS

The most important characteristic of these new materials is a strong correlation among the electrons that makes inapplicable classical schemes based on the band picture. It is necessary to pass from a "single-electron" physics to a "many-electron" physics, where the dominant part will be the correlations among the electrons. Usual perturbation schemes are inadequate and new concepts must be introduced.

The "classical" techniques are based on the hypothesis that the interaction among the electrons is weak and can be treated in the framework of some perturbation scheme. However, as many and many experimental and theoretical studies of highly correlated electron systems have shown with more and more convincing evidence, all these methods are not adequate any more and different approaches must be considered. The main concept that breaks down is the existence of the electrons as particles with some well-defined and intrinsic properties. The presence of interaction modifies the properties of the particles and at a macroscopic level, the level of observation, what are observed are new particles with new peculiar properties entirely determined by the dynamics and by the boundary conditions (i.e. all elements characterizing the physical situation under study). These new objects appear as the final result of the modifications imposed by the interactions on the original particles and contain, by the very beginning, the effects of correlations. Collective behaviors in forms of bound states, resonances, diffused modes and so on emerge as the physical fields. Although some of them are not stable excitations, they give considerable contributions in physical processes, and therefore it is sometimes necessary to promote them to the role of well-defined quasi-particle excitations. The choice of new fundamental particles, whose properties have to be self-consistently determined by dynamics, symmetries and boundary conditions, becomes relevant.

As a simple example, let us consider an atomic system described by the Hamiltonian

$$H = -\mu \sum_{\sigma} \varphi_{\sigma}^{\dagger} \varphi_{\sigma} + V \varphi_{\uparrow}^{\dagger} \varphi_{\downarrow}^{\dagger} \varphi_{\downarrow} \varphi_{\uparrow}$$

(2.1)

$\varphi_{\sigma}$ denotes an Heisenberg electronic field with spin $\sigma = \uparrow, \downarrow$, satisfying canonical anticommutation relations; $\mu$ is the chemical potential and $V$ is the strength of the interaction. This model is exactly solvable in terms of the operators

$$\xi_{\sigma} = \varphi_{\sigma} \varphi_{-\sigma} \varphi_{-\sigma}^{\dagger} \quad \eta_{\sigma} = \varphi_{\sigma} \varphi_{-\sigma}^{\dagger} \varphi_{-\sigma}$$

(2.2)

which are eigenoperators of the Hamiltonian

$$i \frac{\partial}{\partial t} \xi = [\xi, H] = -\mu \xi \quad i \frac{\partial}{\partial t} \eta = [\eta, H] = -(\mu - V)\eta$$

(2.3)

Due to the presence of the interaction, the original electrons $\varphi_\sigma$ are no more observables and new stable elementary excitations, described by the field operators $\xi$ and $\eta$, appear. Due to the $V$-interaction, two sharp features develop in the band structure: the energy level $E = -\mu$ of the bare electron splits in the two levels $E_1 = -\mu$ and $E_2 = V - \mu$. The bare electron reveals itself to be precisely the wrong place to start. A perturbative solution will never give the band splitting.

On the basis of this evidence one can be induced to move the attention from the original fields to the new fields generated by the interaction. The operators describing these excitations, once they have been found, can be written in terms of the original ones and are known as composite operators.

The convenience of developing a formulation to treat composite excitations as fundamental objects has been noticed for the many-body problem of condensed matter physics since long time. Recent years have seen remarkable developments in many-body theory in the form of an assortment of techniques that may be termed composite particle methods. The beginnings of these types of techniques may be traced back to the work of Bogolubov [31], Dancoff [32], Zwanzig [17], Mori [18], Umezawa [33]. The slave boson method, the spectral density approach and the composite operator method (COM) are also along similar lines. This large class of theories is founded on the conviction that an analysis in terms of elementary fields might be inadequate for a system dominated by strong interactions.

All these approaches are very promising because all the different approximation schemes are constructed on the basis of interacting particles: some amount of the interaction is already present in the chosen basis and permits to overcome the problem of finding an appropriate expansion parameter. However, one price must be paid. In general, the composite fields are neither Fermi nor Bose operators, since they do not satisfy canonical (anti)commutation relations, and their properties, because of the inherent definition, must be self-consistently determined. They can only be recognized as fermionic or bosonic operators according to the number, odd or even, of the constituting original electronic fields. This fact makes a tremendous difference with respect to the case of the original electronic operators $\varphi$, which satisfy a canonical algebra.

New techniques of calculus have to be developed in order to treat with composite fields. In developing perturbation calculations where the building blocks are now the propagators of composite fields one cannot use the consolidated scheme: diagrammatic expansions, Wick's theorem and many other techniques are no more valid. The formulation of the Green's function method must be revisited and new frameworks of calculations have to be formulated.

# 3 GREEN'S FUNCTION AND EQUATION OF MOTION FORMALISM

Let us consider a system of $N_e$ interacting Wannier-electrons residing on a Bravais lattice of $N$ sites, spanned by the vectors $\mathbf{R}_i = \mathbf{i}$. For the sake of simplicity, we ignore the presence of magnetic impurities and restrict the analysis to single-band electron models. The generalization of the formalism to more complex systems is straightforward (see for example [34]). In a second quantization scheme this system is described by a certain Hamiltonian

$$H = H[\varphi(i)]$$

(3.1)

describing, in complete generality, the free propagation of the electrons and all the interactions among them and with external fields. $\varphi(i)$ denotes an Heisenberg electronic field [$i = (\mathbf{i}, t)$] satisfying canonical anticommutation relations.

Given the hypothesis that the original fields are not a good basis, we choose a set of composite fields $\{\psi(i)\}$ in terms of which a perturbation scheme will be constructed. Firstly, we choose the set $\psi(i)$ according to the physical properties we want to study. Roughly, the properties of electronic systems can be classified in two large classes: single particle properties, described in terms of fermionic propagators, and response functions, described in terms of bosonic propagators. These two sectors, fermionic and bosonic, are not independent but interplay with each other, and a fully self-consistent solution usually requires that both sectors are simultaneously solved. Once the sector, fermionic or bosonic, has been fixed, we have several criteria for the choice of the new basis. In constructing the composite fields no recipe can be given without thinking to its drawbacks, but many recipes can assure a correct and controlled description of relevant aspects of the dynamics. One can choose: the higher order fields emerging from the equations of motion (in this case the conservation of some spectral moments is assured [35]), the eigenoperators of some relevant interacting terms [36], the eigenoperators of the problem reduced to a small cluster [37].

Let $\psi(i)$ be a n-component field

$$\psi(i) = \begin{pmatrix} \psi_1(i) \\ \vdots \\ \psi_n(i) \end{pmatrix}$$

(3.2)

We do not specify the nature, fermionic or bosonic, of the set $\{\psi(i)\}$. In the case of fermionic operators it is intended that we use the spinorial representation

$$\psi_m(i) = \begin{pmatrix} \psi_{\uparrow_m}(i) \\ \psi_{\downarrow_m}(i) \end{pmatrix} \psi_m^\dagger(i) = \begin{pmatrix} \psi_{\uparrow_m}^\dagger(i) & \psi_{\downarrow_m}^\dagger(i) \end{pmatrix}$$

(3.3)

The dynamics of these operators is governed by the given Hamiltonian $H = H[\{\varphi\}]$ and can be written as

$$i \frac{\partial}{\partial t} \psi(i) = [\psi(i), H] = J(i)$$

(3.4)

In general, this equation cannot be exactly solved and some approximations are necessary. In order to construct approximate solutions one procedure is the following. Let us rewrite the equation of motion as

$$i \frac{\partial}{\partial t} \psi(\mathbf{i}, t) = \sum_{\mathbf{i}} \varepsilon(\mathbf{i}, \mathbf{j}) \psi(\mathbf{j}, t) + \delta J(\mathbf{i}, t)$$

(3.5)

where

$$\varepsilon(\mathbf{i},\mathbf{j}) = \sum_{\mathbf{l}} m(\mathbf{i},\mathbf{l}) I(\mathbf{l},\mathbf{j})^{-1} \qquad \begin{array}{l} I(\mathbf{i},\mathbf{j}) = <[\psi(\mathbf{i},t),\psi^{\dagger}(\mathbf{j},t)]_{\eta}> \\ m(\mathbf{i},\mathbf{j}) = <[J(\mathbf{i},t),\psi^{\dagger}(\mathbf{j},t)]_{\eta}> \end{array} \qquad (3.6)$$

Here $\eta = \pm 1$; usually, it is convenient to take $\eta = 1$ ($\eta = -1$) for a fermionic (bosonic) set $\psi(i)$ (i.e., for a composite field constituted of an odd (even) number of original fields) in order to exploit the canonical anticommutation relations of $\{\psi(i)\}$; but, in principle, both choices are possible. Accordingly, we define

$$[A,B]_{\eta} = \begin{cases} \{A,B\} = AB + BA & for \quad \eta = 1 \\ [A,B] = AB - BA & for \quad \eta = -1 \end{cases} \qquad (3.7)$$

$<\cdots>$ denotes the quantum statistical average over the grand canonical ensemble. Since the components of $\psi(i)$ contain composite operators, the normalization matrix $I(\mathbf{k})$ is not the identity matrix and defines the spectral content of the excitations.

Let us consider the two-time thermodynamic Green's functions (GF) [38]

$$G^{Q}(i,j) = <Q[\psi(i)\,\psi^{\dagger}(j)]> \qquad (3.8)$$

where Q = C (causal), $R$ (retarded), $A$ (advanced). By means of the equation of motion (3.4) we can derive a Dyson equation for composite fields

$$G^{Q}(\mathbf{k},\omega) = G_{0}^{Q}(\mathbf{k},\omega) + G_{0}^{Q}(\mathbf{k},\omega)\Sigma^{Q}(\mathbf{k},\omega)G^{Q}(\mathbf{k},\omega) \qquad (3.9)$$

where $G_{0}^{Q}(\mathbf{k},\omega)$ is the free propagator for composite fields, satisfying the equation

$$[\omega - \varepsilon(\mathbf{k})]G_{0}^{Q}(\mathbf{k},\omega) = I(\mathbf{k}) \qquad (3.10)$$

and $\Sigma^{Q}(\mathbf{k},\omega)$ is the self energy

$$\Sigma^{Q}(\mathbf{k},\omega) = I^{-1}(\mathbf{k})B_{irr}^{Q}(\mathbf{k},\omega)I^{-1}(\mathbf{k}) \qquad (3.11)$$

$B_{irr}^{Q}(\mathbf{k},\omega)$ is the irreducible part of the propagator $B^{Q}(\mathbf{k},\omega) = F.T. <Q[\delta J(i)\delta J^{\dagger}(j)]>$.

We have constructed a generalized perturbative approach designed for formulations using composite fields. Equation (3.9) is a Dyson-like equation and may represents the starting point for a perturbative calculation in terms of the propagator $G_{0}^{Q}(\mathbf{k},\omega)$. Contrarily to the usual perturbation schemes, the calculation of the "free propagator" $G_{0}^{Q}(\mathbf{k},\omega)$ is not an easy task and the next Sections will be dedicated to this problem. Then, the attention will be given to the calculation of the self-energy $\Sigma^{Q}(\mathbf{k},\omega)$, and some approximate methods will be presented. It should be noted that the computation of the two quantities $G_{0}^{Q}(\mathbf{k},\omega)$ and $\Sigma^{Q}(\mathbf{k},\omega)$ are intimately related. The total weight of the self-energy corrections is bounded by the weight of the residual source operator $\delta J(i)$. According to this, it can be made smaller and smaller by increasing the components of the basis $\psi(i)$ [e.g. by including higher-order composite operators appearing in $\delta J(i)$]. The result of such a procedure will be the inclusion in the energy matrix of part of the self-energy as an expansion in terms of coupling constants multiplied by the weights of the newly includes basis operators. In general, the enlargement of the basis leads to a new self-energy with a smaller total weight. However, it is necessary pointing out that this process can be quite cumbersome and the inclusion of fully momentum and frequency dependent self-energy corrections can be necessary to effectively take into account low-energy and virtual processes. According to this, one can chose a reasonable number of components for the basic set and then use another approximation method to evaluate the residual dynamical corrections.

# 4. THE FREE PROPAGATOR $G_0^Q(\mathbf{k}, \omega)$

In this Section we concentrate on the calculation of the Green's functions $G_0^Q(\mathbf{k}, \omega)$ which constitute the building blocks of the perturbation scheme we are trying to formulate. To keep the notation as simple as possible, we will drop the sub index 0 in the definition of $G_0^Q(\mathbf{k}, \omega)$.

One fundamental aspect in a Green's function formulation is the choice of the representation. The knowledge of the Hamiltonian and of the operatorial algebra is not sufficient to completely specify the GF. The GF refer to a specific representation (i.e., to a specific choice of the Hilbert space) and this information must be supplied as a boundary condition to the equations of motion that alone are not sufficient to completely determine the GF. The use of composite operators leads to an enlargement of the Hilbert space by the inclusion of some unphysical states. As a consequence of this, it is difficult to satisfy a priori all the sum rules and, in general, the symmetry properties enjoined by the system under study. In addition, since the representation where the operators are realized has to be dynamically determined, the method clearly requires a process of self-consistency.

From this discussion it is clear that fixing the representation is not an easy task and requires special attention. In the literature the properties of the GF are usually determined by starting from the knowledge of the representation. Owing to the difficulties discussed above we cannot proceed in this way. Therefore, we will derive the general properties of the GF on the basis of the two elements we have: the dynamics, fixed by the choice of the Hamiltonian (3.1), and the algebra, fixed by the choice of the basic set (3.2). The problem of fixing the representation will be considered in Section 7.

Let $\psi(i)$ be a n-component field satisfying linear equations of motion

$$\mathrm{i}\frac{\partial}{\partial t}\,\psi_m(\mathbf{i}, t) = \sum_{\mathbf{j}}\,\sum_{l=1}^{n}\,\varepsilon_{ml}(\mathbf{i}, \mathbf{j})\,\psi_l(\mathbf{j}, t)$$

(4.1)

with the energy matrix $\varepsilon(\mathbf{i}, \mathbf{j})$ defined by (3.6). If the fields $\psi(i)$ are eigenoperators of the total Hamiltonian, the equations of motion (4.1) are exact. If the fields $\psi(i)$ are not eigenoperators of H, the equations are approximated; they correspond to neglecting the residual source operator $\delta J(i)$ in the full equation of motion (3.5) and all the formalism is developed with the intention of using the propagators of these fields as a basis to set up a perturbative scheme of calculations on the ground of the Dyson equation (3.9) derived in the previous Section.

By means of the field equation (4.1), the Fourier transforms of the various Green's functions and of the correlation function $C(i, j) = <\psi(i)\,\psi^\dagger(j)>$ satisfy the following equations

$$[\,\omega - \varepsilon(\mathbf{k})]G^{Q(\eta)}(\mathbf{k}, \omega) = I^{(\eta)}(\mathbf{k})$$
$$[\,\omega - \varepsilon(\mathbf{k})]C(\mathbf{k}, \omega) = 0$$

(4.2)

where the dependence on the parameter $\eta$ has been explicitly introduced. It can be shown that the energy matrix $\varepsilon(\mathbf{k})$ does not depend on the choice of $\eta$. As mentioned in Section 3, the set $\{\psi(i)\}$ can be fermionic or bosonic and the parameter $\eta$ generally takes the value $\eta = 1$ ($\eta = -1$) for a fermionic (bosonic) set $\psi(i)$. The three Green's functions $G^C$, $G^R$ and $G^A$ satisfy the same equation of motion which alone is not sufficient and must be supplemented by other equations. Indeed, the GF are determined by solving a first order differential equation of motion, thereby the GF are given only within an arbitrary constant of integration. The retarded and advanced GF can be completely determined because the factor $\theta[\pm(t_i - t_j)]$ provides the boundary condition: $G^{R,A}(i,j) = 0$ *for* $t_i = t_j \mp \delta$. The determination of the causal GF is not so immediate. In the following we consider the case of finite temperature. For T=0 see Ref. [29].

The most general solution of equation (4.2) is

$$G^{C,R,A,(\eta)}(\mathbf{k},\omega) = \sum_{l=1}^{n} \left\{ P\left( \frac{\sigma^{(l,\eta)}(\mathbf{k})}{\omega - \omega_l(\mathbf{k})} \right) - i\pi\delta[\omega - \omega_l(\mathbf{k})]g^{(l,\eta)C,R,A}(\mathbf{k}) \right\}$$

$$C(\mathbf{k},\omega) = \sum_{l=1}^{n} \delta[\omega - \omega_l(\mathbf{k})]c^{(l)}(\mathbf{k}) \tag{4.3}$$

$g^{(l,\eta)C,R,A}(\mathbf{k})$ and $c^{(l)}(\mathbf{k})$ are momentum functions, not fixed by the equations of motion, to be determined by means of the boundary conditions. $\omega_l(\mathbf{k})$ are the eigenvalues of the matrix $\varepsilon(\mathbf{k})$; $\sigma^{(l,\eta)}(\mathbf{k})$ are the spectral density functions, completely determined by the matrices $\varepsilon(\mathbf{k})$ and $I^{(\eta)}(\mathbf{k})$ as

$$\sigma_{\alpha\beta}^{(l,\eta)}(\mathbf{k}) = \Omega_{\alpha l}(\mathbf{k}) \sum_{\delta} \Omega_{l\delta}^{-1}(\mathbf{k}) I_{\delta\beta}^{(\eta)}(\mathbf{k}) \tag{4.4}$$

where $\Omega(\mathbf{k})$ is the $n \times n$ matrix whose columns are the eigenvectors of the matrix $\varepsilon(\mathbf{k})$. By calculations we obtain [29]:

**Fermionic fields (i.e., $\eta = 1$)**

$$G^{R,A,(+1)}(\mathbf{k},\omega) = \sum_{l=1}^{n} \frac{\sigma^{(l,+1)}(\mathbf{k})}{\omega - \omega_l(\mathbf{k}) \pm i\delta}$$

$$G^{C,(+1)}(\mathbf{k},\omega) = \sum_{l=1}^{n} \sigma^{(l,+1)}(\mathbf{k}) \left[ \frac{1 - f_F(\omega)}{\omega - \omega_l(\mathbf{k}) + i\delta} + \frac{f_F(\omega)}{\omega - \omega_l(\mathbf{k}) - i\delta} \right]$$

$$C(\mathbf{k},\omega) = 2\pi \sum_{l=1}^{n} \delta[\omega - \omega_l(\mathbf{k})]\{1 - f_F[\omega_l(\mathbf{k})]\}\sigma^{(l,+1)}(\mathbf{k}) \tag{4.5}$$

where $f_F(\omega)$ is the Fermi distribution function.

**Bosonic fields (i.e., $\eta = -1$)**

For any given momentum $\mathbf{k}$ we can always write

$$\omega_l(\mathbf{k}) = \begin{cases} = 0 & \text{for } l \in A(\mathbf{k}) \subseteq N = \{1, \ldots n\} \\ \neq 0 & \text{for } l \in B(\mathbf{k}) = N - A(\mathbf{k}) \end{cases}$$

(4.6)

Obviously, $A(\mathbf{k})$ can also be the empty set (i.e., $A(\mathbf{k}) = \varnothing$ and $B(\mathbf{k}) = N$ ).

$$G^{R,A,(-1)}(\mathbf{k}, \omega) = \sum_{l=1}^{n} \frac{\sigma^{(l,-1)}(\mathbf{k})}{\omega - \omega_l(\mathbf{k}) \pm i\delta}$$

$$G^{C,(-1)}(\mathbf{k}, \omega) = -2i\pi \Gamma(\mathbf{k}) \delta(\omega)$$

$$+ \sum_{l \in B(\mathbf{k})} \sigma^{(l,-1)}(\mathbf{k}) \left[ \frac{1 + f_B(\omega)}{\omega - \omega_l(\mathbf{k}) + i\delta} - \frac{f_B(\omega)}{\omega - \omega_l(\mathbf{k}) - i\delta} \right]$$

$$C(\mathbf{k}, \omega) = 2\pi \Gamma(\mathbf{k}) \delta(\omega) + 2\pi \sum_{l \in B(\mathbf{k})} \delta[\omega - \omega_l(\mathbf{k})]\{1 + f_B[\omega_l(\mathbf{k})]\} \sigma^{(l,-1)}(\mathbf{k})$$

(4.7)

where $f_B(\omega)$ is the Bose distribution function. The zero-frequency function (ZFF) $\Gamma(\mathbf{k})$ has been defined as

$$\Gamma(\mathbf{k}) = \frac{1}{2\pi} \sum_{l \in A(\mathbf{k})} c^{(l)}(\mathbf{k}) = \frac{1}{2} \sum_{l \in A(\mathbf{k})} g^{(l,-1)C}(\mathbf{k})$$

(4.8)

and it is left undetermined within the bosonic sector. $\Gamma(\mathbf{k})$ could be computed by considering an anticommutating algebra: remaining in the bosonic sector we make the choice $\eta = 1$ and $\Gamma(\mathbf{k})$ can be calculated by means of the following relations

$$\Gamma(\mathbf{k}) = \frac{1}{2} \sum_{l \in A(\mathbf{k})} \sigma^{(l,+1)}(\mathbf{k}) = \frac{1}{2} \lim_{\omega \to 0} \omega G^{C,(+1)}(\mathbf{k}, \omega)$$

(4.9)

However, the calculation of the $\sigma^{(l,+1)}(\mathbf{k})$ requires the calculation of the normalization matrix $I^{(+1)}(\mathbf{k})$ that, for bosonic fields, generates unknown momentum dependent correlation functions whose determination can be very cumbersome as requires, at least in principle, the self-consistent solution of the integral equations connecting them to the corresponding Green's functions. In practice, also for simple, but anyway composite, bosonic fields the $\Gamma(\mathbf{k})$ remains undetermined and other methods rather than equation (4.9) should be used. Similar methods, like the use of the relaxation function [39], would lead to the same problem.

It is worth pointing out that in the bosonic sector we generally have

$$\sum_{l \in A(\mathbf{k})} \sigma^{(l,-1)}(\mathbf{k}) = 0$$

(4.10)

A situation where $\sum_{l \in A(\mathbf{k})} \sigma^{(l,-1)}(\mathbf{k}) \neq 0$ would lead to the fact that for $l \in A(\mathbf{k})$ the Fourier coefficients $c^{(l)}(\mathbf{k})$ diverge as $[\beta \omega_l(\mathbf{k})]^{-1}$. Since the correlation function in direct space must be finite, at finite temperature this is admissible only in the thermodynamic limit and if the dispersion relation $\omega_l(\mathbf{k})$ is such that the divergence in momentum space is integrable and the corresponding correlation function in real space remains finite. For finite systems and for infinite systems where the divergence is not integrable we must have $\sum_{l \in A(\mathbf{k})} \sigma^{(l,-1)}(\mathbf{k}) = 0$. The calculation of the spectral density matrices $\sigma^{(l,-1)}(\mathbf{k})$ it not a simple dynamical problem, but requires the self-consistent calculation of some expectation values, where the boundary conditions and the choice of the representation play a crucial role. A finite value of $\sum_{l \in A(\mathbf{k})} \sigma^{(l,-1)}(\mathbf{k})$ is generally related to the presence of long-range order and the previous statement is nothing but the Mermin-Wagner theorem [40].

We see that the general structure of the GF is remarkably different according to the statistics. For fermionic composite fields all the Green's functions and correlation functions are completely determined. The zero-frequency function $\Gamma(\mathbf{k})$, defined on the Fermi surface $\omega_l(\mathbf{k}) = \mu$, contributes to the spectral function, is directly related to the spectral density functions $\sigma^{(l,+1)}(\mathbf{k})$ by means of equation (4.9), and its calculation does not require more information. Also, it does not contribute to the imaginary part of the causal GF. For bosonic composite fields the retarded and advanced GF are completely determined, but the causal GF and the correlation function depend on the zero-frequency function $\Gamma(\mathbf{k})$, defined on the surface $\omega_l(\mathbf{k}) = 0$. It is now clear that the causal and retarded (advanced) GF contain different information and that the right procedure of calculation is controlled by the statistics. In particular, in the case of bosonic fields one must start from the causal function and compute the other GF by means of the expressions

$$\mathrm{Re}[G^{R,A(-1)}(\mathbf{k},\omega)] = \mathrm{Re}[G^{C(-1)}(\mathbf{k},\omega)]$$

$$\mathrm{Im}[G^{R,A(-1)}(\mathbf{k},\omega)] = \pm \tanh\left(\frac{\beta\omega}{2}\right)\mathrm{Im}[G^{C(-1)}(\mathbf{k},\omega)]$$

$$C(\mathbf{k},\omega) = -\left[1 + \tanh\left(\frac{\beta\omega}{2}\right)\right]\mathrm{Im}[G^{C(-1)}(\mathbf{k},\omega)]$$

(4.11)

On the contrary, for fermionic fields the right procedure requires first the calculation of the retarded (advanced) function and then computing the other GF by means of the expressions

$$\mathrm{Re}[G^{C(+1)}(\mathbf{k},\omega)] = \mathrm{Re}[G^{R,A(+1)}(\mathbf{k},\omega)]$$

$$\mathrm{Im}[G^{C(+1)}(\mathbf{k},\omega)] = \pm \tanh\left(\frac{\beta\omega}{2}\right)\mathrm{Im}[G^{R,A(+1)}(\mathbf{k},\omega)]$$

$$C(\mathbf{k},\omega) = \mp\left[1 + \tanh\left(\frac{\beta\omega}{2}\right)\right]\mathrm{Im}[G^{R,A(+1)}(\mathbf{k},\omega)]$$

(4.12)

# 5. THE ZERO-FREQUENCY PROBLEM

Given two appropriate operators $A$ and $B$ one can define physical response functions $\chi_{AB}$, called generalized susceptibilities. It was noticed by Kubo [39] that the isolated susceptibility $\chi^I_{AB}(\omega)$, defined for a situation where the system is isolated and the external force is turned on adiabatically, in the limit of zero frequency is in general different from the isothermal susceptibility $\chi^T_{AB}$, defined for a situation where the system is in thermal equilibrium in the presence of a time-independent force. Several years later it was shown [41] that the difference between the two susceptibilities is related to the zero-frequency anomaly exhibited by the bosonic correlation functions, as discussed in the previous Section. Indeed, it can be shown that

$$\chi_{AB}^{T} - \chi_{AB}^{I}(\omega = 0) = \beta \frac{1}{N} \sum_{\mathbf{k}} e^{i\mathbf{k}\cdot(\mathbf{R}_i - \mathbf{R}_j)} \Gamma_{AB}(\mathbf{k}) - \beta < A >< B > \qquad (5.1)$$

Kubo pointed out that the problem of the difference between the two susceptibilities is related to the ergodic property of the system. If the operator $\psi(i)$ has an ergodic dynamics with respect to the Hamiltonian $H$, then the zero-frequency function $\Gamma_{\psi\psi^\dagger}(\mathbf{k})$ must satisfies the following equation

$$\lim_{T \to \infty} \frac{1}{T} \int_0^T < \psi(\mathbf{j},0)\psi^\dagger(\mathbf{j},t) > dt = \frac{1}{N} \sum_{\mathbf{k}} e^{i\mathbf{k}(\mathbf{R}_i - \mathbf{R}_j)} \Gamma_{\psi\psi^\dagger}(\mathbf{k}) = < \psi(\mathbf{i}) >< \psi^\dagger(\mathbf{j}) > \qquad (5.2)$$

If this is true, then the problem of calculating the zero-frequency function is solved and from (5.1) the two generalized susceptibilities $\chi_{AB}^{T}$ and $\chi_{AB}^{I}(0)$ are equal. However, we have not to forget that the condition (5.2) is the same as the standard ergodic requirement only for statistical averages computed in the microcanonical ensemble [39, 42]; in other ensemble it holds only in the thermodynamic limit. Moreover, the condition (5.2) is not satisfied by any integral of motion and, more generally, by any operator that has a diagonal part with respect to the Hamiltonian [43]. This latter consideration clarifies why the ergodic nature of the dynamics of an operator mainly depends on the Hamiltonian which is subject to. It is really remarkable that the zero-frequency constants (ZFC), which are the values of the zero-frequency function $\Gamma(\mathbf{k})$ over the moments for which $A(\mathbf{k}) \neq \varnothing$, are directly related to relevant measurable quantities such as the compressibility, the specific heat, the magnetic susceptibility. According to this, in the case of infinite systems too the correct determination of the zero-frequency constants cannot be considered as an irrelevant issue. In conclusion, Eq. (5.2) generally cannot be used to compute the ZFC and $\Gamma(\mathbf{k})$ has to be computed case by case according to the dynamics and boundary conditions.

The presence of undetermined constants in the bosonic correlation functions is some time known as the zero-frequency anomaly problem. It was first put in evidence in Ref. [44] and then studied by several authors [41, 43, 45-49]. There is a general belief that this problem is of academic interest and in the last years no much attention has been dedicated to it. The main reason is that the response functions, the experimentally observed quantities, are given by retarded bosonic GF which, as we have shown, formally do not depend on the zero-frequency constants, which are, therefore, considered of no physical interest. The general attitude [39, 45] is to believe that in macroscopic real systems at equilibrium at temperature T, the fluctuations are very small and the interaction between the system and the reservoir would introduce an irreversible relaxation and decouple the correlation functions. Then, as suggested in Ref. [45], the zero-frequency constants should be always determined by requiring the ergodicity and therefore fixed by means of Eq. (5.2). This procedure is some how an artifice and may lead to serious problems because it might break the internal self-consistency of the entire formulation. The fact that the retarded GF do not depend on the zero-frequency constant is true only for noninteracting systems. In general, for interacting systems the retarded GF do depend on the ZFC. To understand this we must recall that in the equations of motion of all the GF appears an inhomogeneous term, the normalization matrix $I(\mathbf{i}, \mathbf{j}) = < [\psi(\mathbf{i}, t), \psi^\dagger(\mathbf{j}, t)]_\eta >$. This quantity is expressed in terms of various correlation functions, depending on the algebra of the set $\{\psi(i)\}$, of fermionic and bosonic nature, to be determined in a self-consistent way. Since the bosonic correlation functions depend on the ZFC, the normalization function and therefore all the GF do depend on the ZFC. These quantities cannot be fixed in an arbitrary way, but they must be calculated in order to keep the internal self-consistency of the global formulation.

## 6. ARE THE GREEN'S FUNCTIONS FULLY DETERMINED?

By means of the equations of motion and by using the boundary conditions related to the definitions of the various Green's functions we have been able to derive explicit expressions for these latter [cfr. (4.5) and (4.7)]. However, these expressions can only determine the functional dependence; the knowledge of the GF is not fully achieved yet. The reason is that the algebra of the field $\psi(i)$ is not canonical. As a consequence, the inhomogeneous terms $I^{(\eta)}(\mathbf{k})$ in the equations of motion (4.2) and the energy matrix $\varepsilon(\mathbf{k})$ contain some unknown static correlation functions, correlators, that have to be self-consistently calculated. These functions can be both of fermionic and bosonic nature and usually one needs to study more sectors at the same time. Furthermore, these correlation functions are expectation values of higher-order operators not belonging to the chosen basis $\{\psi(i)\}$. This is the most serious problem! In order to calculate these correlators one should enlarge the basis by including the new higher-order operators and repeat the scheme of calculation. It is clear that the calculation of the new matrices $I^{(\eta)}(\mathbf{k})$ and $\varepsilon(\mathbf{k})$ will lead to new correlators and new higher-order field operators will appear. In general the process might not converge, or a huge number of basic operators will be needed. In addiction to this problem, in the case of bosonic fields, there is the presence of the zero-frequency functions $\Gamma(\mathbf{k})$ whose determination is not easy at all. The self-consistent calculation of the unknown correlators and zero-frequency functions must be performed in order to completely determine the GF. It is important to remark that the entire process of self-consistency will affect all the GF at the same time and, therefore, all the physical properties of the system. For instance, as noticed in the previous Section, although the retarded GF do not explicitly depend on the ZFC, there is an implicit dependence through the internal self-consistent parameters. A self-consistent scheme of calculations for the various GF will be given in the next Section.

## 7. A SELF-CONSISTENT SCHEME

In the approximation scheme we are proposing, an essential element is the knowledge of the free propagators $G_0^Q(\mathbf{k}, \omega)$. These quantities have been largely studied in Section 5 and the explicit expressions have been obtained. However, three serious problems arise with the study of these functions:

(a) the calculation of some parameters expressed as correlation functions of field operators not belonging the chosen basis;

(b) the appearance of some zero-frequency constants (ZFC) and their determination;

(c) the problem of fixing the representation where the Green's functions are formulated.

In most of the approaches found in the literature the solution of the previous problems is the following.

(a) In order to determine the unknown parameters several methods (arbitrary ansatz, decoupling schemes, use of the equation of motion) have been considered in the context of different approaches (Hubbard I approximation [3], Roth's method [50], projection method [2], spectral density approach [4, 5]). As shown in Ref. 28 in the context of the Hubbard model, these procedures lead to a series of unpleasant results: several sum rules and the particle-hole symmetry are violated, there is no presence of a Mott transition, all local quantities strongly disagree with the results of the numerical simulation.

(b) The ZFC are usually fixed by requiring the ergodicity of the dynamics of the relative operators. This is clearly a very strong assumption. There are many examples where the zero-frequency constants do not assume their ergodic value: if we would force the ZFC to assume it, this choice leads to wrong results. In general, these quantities must be calculated case by case.

(c) The knowledge of the Hamiltonian and of the operatorial algebra is not sufficient to completely specify the GF. The GF refer to a specific representation (i.e., to a specific choice of the Hilbert space) and this information must be supplied to the equations of motion that alone are not sufficient to completely determine the GF. The construction of the Hilbert space where the GF are realized is not an easy task and is usually ignored. The use of composite operators leads to an enlargement of the Hilbert space by the inclusion of some unphysical states. As a consequence of this, it is difficult to satisfy a priori all the sum rules and, in general, the symmetry properties enjoyed by the system under study.

In the composite operator method (COM) the three problems are not considered separately but they are all connected in one self-consistent scheme. The main idea is that fixing the values of the unknown parameters and of the ZFC implies to put some constraints on the representation where the GF are realized. As the determination of this representation is not arbitrary, it is clear that there is no freedom in fixing these quantities. They must assume values compatible with the dynamics and with the right representation. Which is the right representation? This is a very hard question to answer.

From the algebra it is possible to derive several relations among the operators. We will call Pauli constraints (PC) all possible relations among the operators dictated by the algebra. This set of relations valid at microscopic level must be satisfied also at macroscopic level, when expectations values are considered. In general, the correlation functions calculated by means of the equation of motion, as shown in Section 4, without having specified the representation, do not satisfy the relations called by the algebra. To see this, let us consider as an example the correlation function $C_{\xi\eta^\dagger}(i,j) = \langle \xi(i)\,\eta^\dagger(j) \rangle$, where $\xi(i)$ and $\eta(i)$ are the Hubbard operators defined by Eq. (2.2). Owing to the fact that the algebra of these operators is not canonical, the correlation function $C_{\xi\eta^\dagger}(i,j)$ will depend on a set of parameters $\{p_1, p_2, \ldots p_m\}$, not known a priori, which must be calculated by some appropriate methods. By means of the Pauli principle, the operators $\xi(i)$ and $\eta(i)$ satisfy the relation $\xi(i)\eta^\dagger(i) = 0$. However, when we consider the expectation value it is clear that the relation

$$\langle \xi(i)\,\eta^\dagger(i) \rangle = 0 \qquad (7.1)$$

will be satisfied only when the parameters $\{p_1, p_2, \ldots p_m\}$ will take appropriate values. For any other values the relation (7.1) will be violated. It is then evident that there is no freedom in determining the parameters $\{p_1, p_2, \ldots p_m\}$. If (7.1) is not satisfied, it is clear that in the Hilbert space we are picking up states of the type $\left| i(\uparrow), i(\uparrow) \right\rangle$, which are incompatible with the Pauli principle and must be eliminated.

We also note that, in general, the Hamiltonian has some symmetry properties (i.e. rotational invariance in coordinate and spin space, phase invariance, gauge invariance,.......). These symmetries generate a set of relations among the matrix elements: the Ward-Takahashi identities [51] (WT).

Now, certainly the right representation must be the one where all relations among the operators satisfy the conservation laws present in the theory when expectation values are taken (i.e., where all the PC and WT are preserved). Then, we impose these conditions and obtain a set of self-consistent equations that will fix the unknown correlators, the ZFC and the right representation at the same time. Several equations can be written down, according to the different symmetries we want to preserve. A large class of self-consistent equations is given by the following equation

$$\langle \psi(i)\,\psi^\dagger(i) \rangle = \frac{1}{N}\sum_{\mathbf{k}}\frac{1}{2\pi}\int_{-\infty}^{+\infty}d\omega\, C_{\psi\psi^\dagger}(\mathbf{k},\omega) \qquad (7.2)$$

where the l.h.s. is fixed by the PC, the WT and the boundary conditions compatible with the phase under investigation and in the r.h.s. the correlation function $C_{\psi\psi^\dagger}(\mathbf{k},\omega)$ is computed by means of the equation of motion, as illustrated in Section 4. Equations (7.2) generate a set of self-consistent equations which determine the unknown parameters (i.e., ZFC and unknown correlators) and, consequently, the proper representation, avoiding the problem of uncontrolled and uncontrollable decoupling. Condition (7.2) can be considered as a generalization, to the case of composite fields, of the equation that, in the non-interacting case, fixes the way of counting the particles per site, according to the algebra, by determining the chemical potential.

Another important relation, that will be largely used in the applications, is the requirement of time translational invariance which leads to the condition that the m-matrix, defined by Eq. (1.3.14), must satisfy the following relation:

$$m_{ab}(\mathbf{k}) = \left(m_{ba}(\mathbf{k})\right)* \qquad (7.3)$$

This is a particular case of a more general condition on the spectral moments [35]. It can be shown that if (7.3) is violated, then states with a negative norm are included in the Hilbert space.

It should be noted that the number of constraints generated by Eqs. (7.2) and (7.3) can be different from the number of unknown parameters. Generally, the coincidence of these two numbers signals that the chosen basic set gives a reasonable description of the dynamics.

It is worth noting that by means of Eqs. (7.2) is often possible to close one sector (i.e., fermionic, spin, charge, pair, ...) at a time without resorting to the opening of all or many of them simultaneously. Obviously, this occurrence enormously facilitates the calculations.

# 8. THE DYSON EQUATION

The generalized Dyson equation (3.9) is an exact equation and permits, in principle, once the normalization matrix $I(\mathbf{i}, \mathbf{j})$, the m-matrix $m(\mathbf{i}, \mathbf{j})$ and the propagator $B(i, j)$ are known, in the framework of the self-consistent scheme outlined in Section 7, the calculation of the various Green's functions. However, for most of the physical systems of interest the calculation of the propagator $B(i, j)$ is a very difficult task and some approximations are needed. Various approximate schemes have been proposed.

The simplest approximation is based on completely neglecting the dynamical part $\Sigma(\mathbf{k}, \omega)$. This approximation is largely used in the literature [2-5, 18-21, 23-29, 50, 52-58] and is called pole approximation. This approximation consists in retaining that one can neglect finite life-time effects paying attention to the choice of a proper extended operatorial basis, with respect to which the self-energy corrections have a small total weight. Indeed, the total weight of the corrections is bounded by the thermal average involving the residual source $\delta J(i)$. It is worth noting [35] that the n-pole structure of the various GF corresponds to a Dyson-like equation

$$G_{ab}^{Q}(\mathbf{k}, \omega) = \frac{I_{ab}(\mathbf{k})1}{\omega - \Sigma_{ab}^{Q}(\mathbf{k}, \omega)} \tag{8.1}$$

where the self-energy components $\Sigma_{ab}^{Q}(\mathbf{k}, \omega)$ have a (n-1)-pole structure. A theorem concerning the conservation of the spectral moments $M^{(p)}(\mathbf{k}) = F.T.\left\langle \left[ \left( i\partial / \partial t \right)^{p} \psi(\mathbf{i}, t), \psi^{\dagger}(\mathbf{j}, t) \right]_{\eta} \right\rangle$ can be assessed [35].

**Theorem:** Consider a Hamiltonian $H$ and choose a set of composite fields $\{ \psi_{l}, l = 1, \cdots n \}$. If the subset $\{ \psi_{l}, l = 1, \cdots n - 1 \}$ is chosen so that

$$i\frac{\partial}{\partial t} \psi_{l}(i) = [ \psi_{l}(i), H ] = \sum_{p=1}^{l+1} \gamma_{lp}(-i\nabla) \psi_{p}(i) \qquad \text{[for } 1 \leq l \leq n - 1] \tag{8.2}$$

then the first $2(n - l + 1)$ spectral moments for the field $\psi_{l}(i)$ $(1 \leq l \leq n - 1)$ are conserved.

In other words, the conservation of the first $2(n - l + 1)$ spectral moments is automatically assured if we construct a multiplet whereby, at any stage, the sources rule what should enter as a new operator.

As a corollary, this theorem shows that the n-pole approximation is equivalent to the spectral density approach [4, 5] when the specific choice (8.2) for the basis is considered. However, it is important to remark that the choice (8.2) suffers from severe limitations. For several systems, for example for a multi-orbital model, a basis diagonalizing the atomic problem could be more appropriate than the one coming from the equations of motion [23, 34]. In some other cases, by choosing the appropriate field it is possible to catch the low-energy physics of the system [59]. This is unfeasible through a finite sum of spectral moments as we would need an increasingly large number of them to describe lower and lower energy scales.

In order to go beyond the n-pole approximation one needs to take into account self-energy corrections by developing some methods to calculate the effects of $\Sigma(\mathbf{k}, \omega)$. Various approximate schemes have been proposed. We mention some of them.

**Born approximation**

In the self-consistent Born approximation (SCBA), or non-crossing approximation, the many-particle Green's functions, appearing in the expression of $\Sigma(\mathbf{k}, \omega)$ [see (3.11)], are calculated by assuming that the fermionic and bosonic modes propagate independently. In order to illustrate the approximation, let us consider the case where the basic set $\{\psi(i)\}$ is of a fermionic type. Then, typically we have to calculate GF of the form

$$H^R(i,j) = < R[B(i)F(i)F^\dagger(j)B^\dagger(j)]>$$ (8.3)

where $F(i)$ and $B(i)$ are fermionic and bosonic fields, respectively. By means of (4.12) we can write

$$H^R(\mathbf{k}, \omega) = -\frac{1}{\pi} \int_{-\infty}^{+\infty} d\omega' \frac{1}{\omega - \omega' + i\varepsilon} \coth \frac{\beta\omega'}{2} \operatorname{Im}[H^c(\mathbf{k}, \omega')]$$ (8.4)

where $H^C(i,j) = < T[B(i)F(i)F^\dagger(j)B^\dagger(j)]>$ is the causal function. In the SCBA we approximate

$$H^C(i,j) \approx f^c(i,j)b^c(i,j)$$ (8.5)

where $f^C(i,j) = < T[F(i)F^\dagger(j)]>$ and $b^C(i,j) = < T[B(i)B^\dagger(j)]>$. Approximation (8.5) has been used in many works (as an example see 60-62) .By assuming that the system is ergodic we can use the spectral representation to obtain

$$H^R(\mathbf{k}, \omega) = \frac{1}{\pi} \int_{-\infty}^{+\infty} d\omega' \frac{1}{\omega - \omega' + i\delta} \frac{a^d}{(2\pi)^{d+1}} \int_{\Omega_B} d^d p d\Omega \operatorname{Im}[f^R(\mathbf{p}, \Omega)]$$

$$\operatorname{Im}[b^R(\mathbf{k} - \mathbf{p}, \omega' - \Omega)][\tanh \frac{\beta\Omega}{2} + \coth \frac{\beta(\omega' - \Omega)}{2}]$$ (8.6)

**Two-site resolvent approach**

In this scheme [26, 27] the dynamical part $\Sigma(\mathbf{k}, \omega)$ of the self-energy is estimated by a two-site approximation in combined use with the resolvent method [6]. Let us approximate the higher order propagator as

$$B^Q(\mathbf{k}, \omega) = F.T. < Q[\delta J(i)\delta J^\dagger(j)]> \approx B_0^Q(\omega) + \alpha(\mathbf{k})B_1^Q(\omega)$$ (8.7)

where $B_0^Q(\omega)$ is related to level transitions on equal site, while $B_1^Q(\omega)$ is related to transitions across the two sites. The Green's function (3.8) takes the form

$$G^Q(\mathbf{k}, \omega) = \frac{1}{\omega - \varepsilon(\mathbf{k}) + t^2 V(\omega)\alpha(\mathbf{k})} I(\mathbf{k})$$ (8.8)

where $V(\omega)$ has to be calculated from the definition (3.11) by making use of (8.7). This approximation has been applied to the study of the t-J [26] and Hubbard [27, 63] models. It has been shown that the approach produces most of the features seen in the numerical simulation as well as the features of spectral distributions near the metal-insulator transition.

# 9. CONCLUSIONS

I have illustrated a formalism for the study of highly correlated electronic systems, based on two main concepts: propagators of composite operators as building blocks of a perturbation calculation; use of algebra constraints to fix the representation of the GF in order to maintain the algebraic and symmetry properties. The outline of the method can be so schematized:

(i) Given a certain Hamiltonian expressed in terms of electronic fields, one chooses a set $\{\psi(i)\}$ of composite operators.

(ii) A generalized Dyson equation is derived

$$G(\mathbf{k}, \omega) = G^{(0)}(\mathbf{k}, \omega) + G^{(0)}(\mathbf{k}, \omega) \Sigma(\mathbf{k}, \omega) G(\mathbf{k}, \omega)$$

where $G(\mathbf{k}, \omega)$ is the complete GF and $G^{(0)}(\mathbf{k}, \omega)$ is the "free" propagator obtained by linearizing the dynamics of the composite fields through a projection on the basis itself.

(iii) The functional dependence of $G^{(0)}(\mathbf{k}, \omega)$ in terms of internal parameters (ZFC and correlators) is determined.

(iv) The internal parameters are determined by a set of self-consistent equations which restore the algebra constraints and the symmetry properties of the Hamiltonian.

(v) An approximation is chosen for the determination of the dynamical self-energy $\Sigma(\mathbf{k}, \omega)$.

During the last years this formulation has been applied to the study of several systems: Hubbard, t-t'-U, t-J, p-d, double exchange, Kondo, Anderson, Heisenberg models. A systematic comparison with the results of numerical simulation has been carried out. The interested reader may refer to the works cited in the bibliography and references therein.

# ACKNOWLEDGMENTS

This manuscript summarizes the research work we have done at the University of Salerno in the last eighth years. During this period, the collaborations with Hideki Matsumoto and Adolfo Avella have been of essential importance. They have brought decisive and vital contributions and I take this opportunity to express to them my deepest gratitude. I also wish to mention the numerous graduate students and post-doc fellows that during these years took part in various phases of the work giving significative contributions: A.M. Allega, M. Bak, T. Di Matteo, S.-S. Feng, V. Fiorentino, S. Krivenko, S. Marra, R. Münzner, S. Odashima, V. Oudovenko, N. Perkins, T. Saikawa, M.M. Sanchez, V. Turkowski, D. Villani, E. Zasinas. I also wish to acknowledge several international collaborations: A.F. Barabanov, F.D. Buzatu, R. Hayn, N.M. Plakida, L. Siurakshina, R. Sridhar, M. Tachiki, V.Yu. Yushankhai.